\newcounter{myctr}
\def\myitem{\refstepcounter{myctr}\bibfont\noindent\ifnum\themyctr>9\else\phantom{0}\fi\hangindent17pt\themyctr.\enskip}

\documentclass{ws-ijqi_tama}
\usepackage{hyperref}
\usepackage[super,sort,compress]{cite}

\usepackage{amsmath,amssymb,amsfonts,bbm,graphicx,hyperref,color,times}
\usepackage{subfigure}
\usepackage[english]{babel}

\newcommand{\bra}[1]{\ensuremath{\langle #1|}}
\newcommand{\ket}[1]{\ensuremath{|#1 \rangle}}
\newcommand{\braket}[2]{\ensuremath{\langle #1| #2 \rangle}}
\newcommand{\ketbra}[2]{\ensuremath{| #1 \rangle\hspace{-2pt} \langle #2 |}}

\newcommand{\eref}[1]{(\ref{#1})}
\newcommand{\fref}[1]{figure \ref{#1}}
\newcommand{\Fref}[1]{Figure \ref{#1}}

\newcommand{\llrr}[1]{\ensuremath{\left( #1\right)}}

\newcommand{\llrrb}[1]{\ensuremath{\left \{ #1\right \} }}

\DeclareMathOperator{\diag}{\hbox{diag}}

\begin{document}

%%%%%%%%%%%%%%%%%%%%% Publisher's Area please ignore %%%%%%%%%%%%%%
\catchline{}{}{}{}{}
%%%%%%%%%%%%%%%%%%%%%%%%%%%%%%%%%%%%%%%%%%%%%%%%%%%%%%%%%%%%%%%%%%%

\title{DEPHASING ASSISTED TRANSPORT\\ ON A BIOMIMETIC RING STRUCTURE}

\author{DARIO TAMASCELLI}
\address{Quantum Technology Lab, Dipartimento di Fisica, 
Universit\`a degli Studi di Milano, I-20133 Milano, Italy\\
%Institut f\"ur Theoretische Physik \&  IQST, Albert-Einstein-Allee 11, Universit\"at Ulm, Germany\\
dario.tamascelli@unimi.it}
\author{ALESSIA SEGATI}
\address{Dipartimento di Fisica, 
Universit\`a degli Studi di Milano, I-20133 Milano, Italy}
\author{STEFANO OLIVARES}
\address{Quantum Technology Lab, Dipartimento di Fisica, 
Universit\`a degli Studi di Milano, I-20133 Milano, Italy\\
INFN, Sezione di Milano, I-20133 Milano, Italy\\
stefano.olivares@fisica.unimi.it}

\maketitle

%\begin{history}
%\received{Day Month Year}
%\revised{Day Month Year}
%%\accepted{Day Month Year}
%%\comby{(xxxxxxxxxx)}
%\end{history}

%%%%%%%%%%%%%%%%%%%%%%%%%%%%%%%%%%%%%%

\begin{abstract}
We address two-level systems arranged in ring configurations affected by static disorder. In particular we investigate the
role of dephasing in the transport of an excitation along the ring.
%We investigate the transport of an excitation along a ring of two-level systems and show that dephasing enhances the transfer capability of the system in the presence of static disorder. 
We compare the efficiency of the transfer process on isotropic rings and on biomimetic rings modelled
according to the structure of light-harvesting complexes. Our analysis provides a simple but clear and
interesting example of how an interplay between the coherent dynamics of the system and
the incoherent action of the environment can enhance the transfer capabilities of disordered
lattices. 
\end{abstract}


%\tableofcontents  % optional

\markboth{D. Tamascelli, A. Segati and S. Olivares}
{Dephasing assisted transport on a biomimetic ring structure}

%%%%%%%%%%%%%%%%%%%%%%%%%%%%%%%%%%%%%%

\section{Introduction}
Quantum coherent evolution can provide a substantial advantage with respect to classical random walks for the task of transferring energy or information between selected vertices of a graph
\cite{childs02, childs03, christandl05,tama16}. The latter, in turn, can be used to model the different parties of a composite system, such as, for example, a communication network or a biological complex \cite{bose07,mulken11}.
%Quantum walk on graphs have therefore received much attention in the last decade. 
Recent experiments have revealed the presence of quantum coherence in the Exciton Energy
Transfer processes (EET) that occur on the photosyntetic membranes of purple bacteria \cite{Engel2007,Lee2007,panit2010}. This
observations triggered quite a large amount of research aimed to understanding whether Nature itself is exploiting quantum
coherence as a resource in EET, which is well known to be a remarkably efficient biological
processes \cite{sension2007,oyala2008,plenio2008,caruso09,sarovan2010,smyth2012,sangwoo12}. The
understanding of the fundamental mechanism that allows for such efficiency can provide design
principles for synthetic photovoltaic devices \cite{scholes2011}. 

Coherence, on the other side, is usually washed away by the unavoidable interaction of the system with the surrounding
environment which is therefore expected to detriment the EET efficiency. Recent research, however,
showed that there are situations where the presence of a
certain amount of decoherence can actually be beneficial for transport \cite{plenio2008,rebentrost08,mohseni08,caruso09,contreras14}. In the presence of static
disorder, for example, dephasing can help delocalizing an excitation that would otherwise suffer from  Anderson localization. 
This phenomenon is known as Dephasing Assisted Transport (DAT).\cite{plenio2008,caruso09}

In this paper we investigate excitation transfer on a ring, i.e. $N$-cycle graph, in the
presence of disorder and decoherence. We consider $N$-cycles graphs with isotropic and
alternate nearest-neighbour coupling. The latter configuration resembles the structure of Light
Harvesting complexes of type 1, called LH1 (see Ref.s~\citen{Niwa2014,felipe15}), and
2, LH2 (see Ref.~\citen{zhao14}) that can be found on the photosynthetic
membrane of purple bacteria.  The aim of the paper is to provide simple but interesting examples of the possible
role of dephasing in transport over disordered $N$-cycle lattices. In particular we show that dephasing
can assists transport on both types of rings, when the on-site energies
of the cycles are affected by static disorder. We moreover find that alternate couplings do not
offer any functional advantage with respect to excitation transfer between two opposite sites of
the $N$-cycle.

The paper is organized as follows. In Section 1 we introduce and characterize the ring
models and define the the adopted EET efficiency measure. In Section 2 we define the disorder
and dephasing and study  their relative effects on EET. Section 3 presents the analysis of the
conbined effects of disorder and dephasing. The last section is devoted to conclusion and
outlook.%

\section{The model}
The quantum system we consider in this paper is a collection of $N$ two-level systems (TLSs)
arranged in a circular structure. Each TLS $j$, $j=1,2,\ldots,N$, can be in its
ground, $\ket{0}_j$, or excited, $\ket{1}_j$ state. We restrict our analysis to the
relevant single excitation case \cite{felipe11}, where only one of the TLSs is excited while the others
are all in the ground state. Defined $\ket{j} = \otimes_{n=1}^N \ket{\delta_{j,n}}_n$, the set $\{
    \ket{j} \}_{j=1}^N$ is a basis for the Hilbert space of considered states. 
In this basis the system Hamiltonian reads:
%%%%%%
\begin{eqnarray}
    H = \sum_{j=1}^{N} J(j)( \ketbra{j+1}{j}+\ketbra{j}{j+1}) + \varepsilon(j) \ketbra{j}{j},
    \label{eq:genHam}
\end{eqnarray}
%%%%
with the condition $\ket{N+1}=\ket{1}$ accounting for the closed boundary conditions. The
coefficients $J(j)$ represent the coupling strength between nearest-neighbour sites whereas
$\varepsilon(j)$ are the on site energy difference between the ground and excited state of the $j$-th TLS.
\\In what follows we will study the transfer of the single excitation for two different parametrizations of the ring
Hamiltonian $H$. In the first one we consider isotropic couplings $J(j)=J$ and equal on-site energies
$\varepsilon(j) = \varepsilon$ (without loss of generality we can set, in this case, $\varepsilon=0$). 
The Hamiltonian corresponding to this parametrization reduces to:
\begin{align}
    H_{\rm R} = J \sum_{j=1}^N \ketbra{j+1}{j} + \ketbra{j}{j+1}. 
    \label{eq:hamR}
\end{align}
 The second parametrization is chosen as to resemble the configuration of the LH1 complex of  purple photosynthetic bacteria (\emph{Rhodospirillum rubrum}) \cite{felipe15}. 
 This complex is formed by 32 BChl molecules, bound to 16 $\alpha$ and $\beta$ polypeptides as $\alpha\beta(\text{BChl)}_2$ subunits 
 organized in a ring geometry. 
 The $\alpha$ and $\beta$ polypeptides of nearest-neighbour subunits are interacting with each other with \emph{inter-dimer} coupling 
 strength $J_2=377 \mbox{cm}^{-1}$ 
 whereas  the $\alpha \beta$ BChls in the same subunit are coupled with \emph{intra-dimer} 
 coupling coefficient $J_1=600 \mbox{cm}^{-1}$.
 While the on-site energies $\varepsilon_\alpha$ and $\varepsilon_\beta$ of the LH1
 for the $\alpha$ and $\beta$ polypeptides are slighty different
 ($\varepsilon_\alpha/\varepsilon_\beta = 0.999$), we assume, for the sake of simplicity,
 $\varepsilon_\alpha = \varepsilon_\beta$. The Hamiltonian for this \emph{biomimetic} configuration of the ring can be written as:
\begin{align}
    H_{\rm LH1} &= \sum_{j=0}^{N/2} J_1 (\ketbra{2j+2}{2j+1} + \ketbra{2j+1}{2j+2}) \nonumber \\
    &\hspace{2cm}+ \sum_{j=1}^{N/2} J_2 (\ketbra{2j+1}{2j} + \ketbra{2j}{2j+1})
    \label{eq:hamLH1},
\end{align}
where we have omitted the dynamically irrelevant on-site energies $\varepsilon_j$, and $N=32$. As to allow for a direct comparison between the two ring 
configurations, we will set for the system described by \eref{eq:hamR} $N=32$ and $J=(J_1+J_2)/2$. 

\begin{figure}[h]
\centering
\includegraphics[width=0.50\columnwidth]{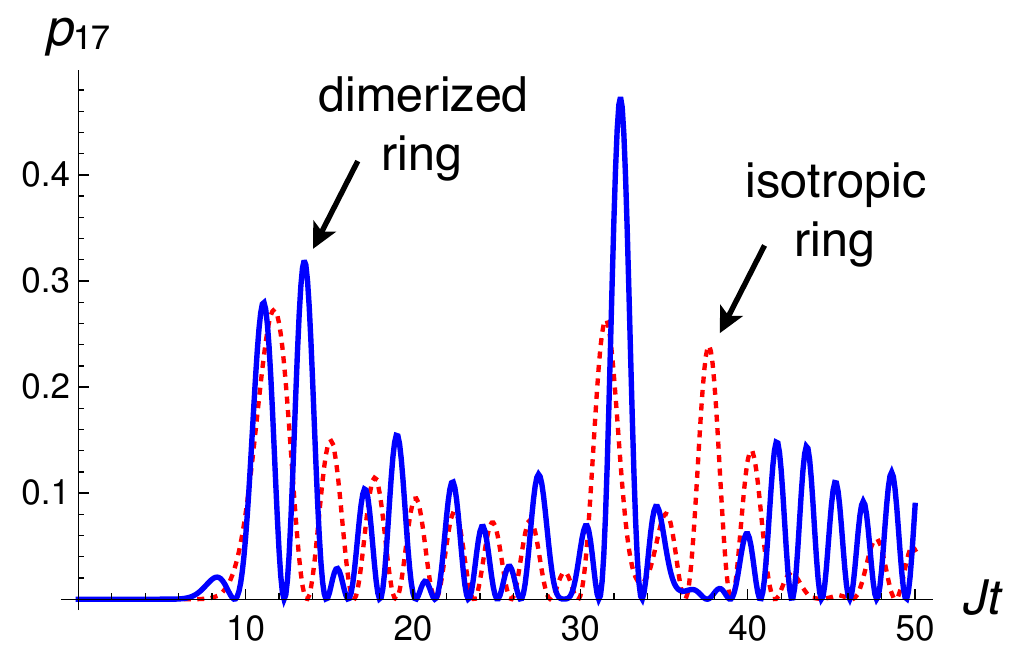}
\caption{The transfer probability $p_{17}(t)=|\braket{17}{\psi^{\rm R/LH}}{(t)}|^2$ as a function of time, with
$\ket{\psi^{\rm R/LH}(t)}=e^{-i t H_{\rm R}/H_{\rm LH}}\ket{1}$. Dotted (red) line:
isotropic ring with $J=(J_1+J_2)/2$. Solid (blue) line: dimerized ring. $J_1=600$, $J_2=377$,
$J=(J_1+J_2)/2$.}
\label{fig:gP17}
\end{figure}

The spectra of $H_{\rm R}$ and $H_{\rm LH1}$ can both be analytically computed.
For the isotropic case, the eigenvectors of $H_{\rm R}$ are given by the Fourier basis
\begin{align}
    \ket{e_k^{\rm R}} &= \frac{1}{\sqrt{N}} \sum_{j=1}^N e^{i \frac{2 \pi k}{N} j  } \ket{j},\quad
    k=1,2,\ldots,N
    \label{eq:eigenR}
\end{align}
with corresponding eigenvalues $e_k^{\rm R} = -2 J \cos \llrr{\frac{2 \pi k}{N}}$. To solve the
eigenvalue/vector problem for the LH1 dimerized Hamiltonian \eref{eq:hamLH1}, it is instead
expedient to use a relabelling of the $j$-th TLS of the ring with the couple $\llrr{n=\lfloor j/2 \rfloor,s=(j \mod
2) +1}$ and to define the Fourier basis
\begin{align}
    \ket{k,s} =\frac{1}{\sqrt{K}} \sum_{n=0}^{K-1} e^{i \frac{2 \pi}{K} k n } \ket{n,s}
    \label{eq:fourierBasisDimer}.
\end{align}
The Hamiltonian \eref{eq:hamLH1} assumes, in this basis, a
block-diagonal form whose $2 \times 2$ matrices on the diagonal are:
 \[
   h_k=\llrr{ 
       \begin{array}{c c}
           0 & J_1 + e^{-i \alpha_k} J_2 \\
           J_1 + e^{i \alpha_k} J_2& 0
    \end{array}
},
\]
with $\alpha_k = 2 \pi k /K$ and $K=N/2$; each $h_k$ has eigenvalues $\pm  e_k^{\rm LH}$, with
$ e_k^{\rm LH}=  J_2 \sqrt{\beta^2+1+2  \beta \cos \alpha_k}$,
where $\beta = J_1/J_2$ indicates the degree of dimerization and is diagonalized by the transformation
 \[
     U_k=\frac{1}{\sqrt{2}}\llrr{ 
       \begin{array}{c c}
           \eta_k & -\eta_k \\
           1& 1
    \end{array}
},
\]
with  $\eta_k = (\beta + e^{-i \alpha_k})/e_k^{\rm LH}$.
\begin{figure}[h]
\centering
\includegraphics[width=0.50\columnwidth]{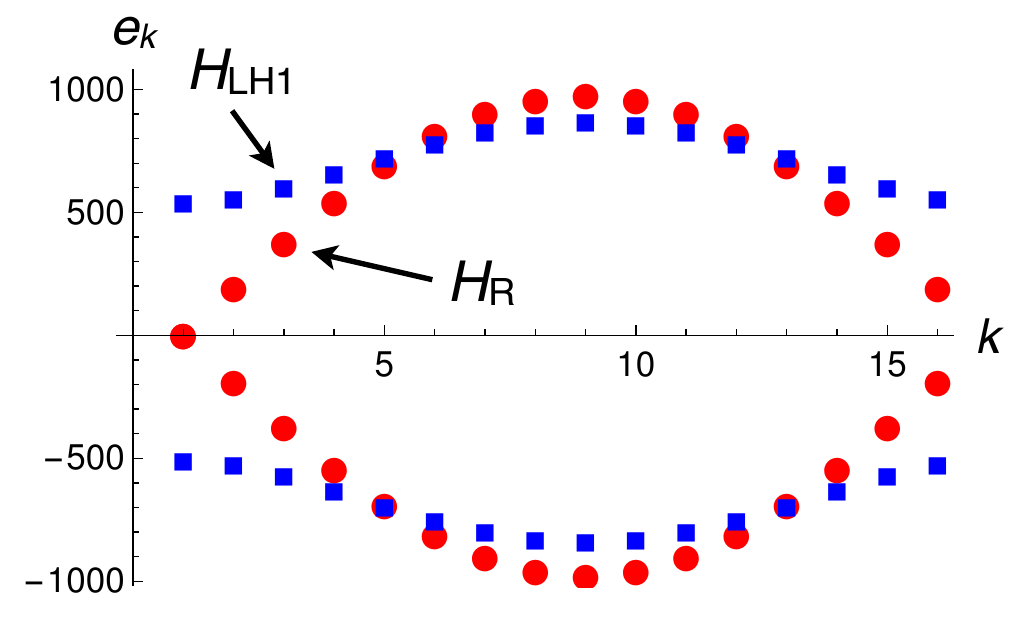}
\caption{The spectra of $H_{\rm R}$ (circles) and $H_{\rm LH1}$ (squares) for the parametrization given in the
main text.}
\label{fig:gSpectrum}
\end{figure}
\Fref{fig:gSpectrum} shows the spectra of the isotropic (circles) and the biomimetic, or \emph{dimerized}, (squares) 
rings: in both cases the spectra have two bands. In the presence of alternate couplings $J_1,J_2$,
however, the spectrum shows an energy gap $2 J_2 \sqrt{1+\beta^2}$, i.e. directly proportional to the degree
of dimerization $\beta$.
In order to study the transport through the rings we assume that the excitation is initially localized on the 
TLS $j=1$, namely $\ket{\psi(0)} = \ket{1}$, and evaluate the probability of finding the excitation at the opposite site $j=17$ 
at time $t$, i.e. $|\braket{17}{\psi(t)}|^2 \stackrel{\text{def}}{\equiv}p_{17}(t)$. The result is shown in \fref{fig:gP17}.
Due to the symmetric couplings
of site $j=1$ to its nearest-neighbours, in the case of an isotropic ring the initial condition
separates in two wavefronts of equal amplitudes that propagate at the same rate along the ring  in opposite directions and
arrive ``simultaneously'' at the opposite site. In the presence of alternate couplings, the broken
symmetry is made evident by the two separated peaks (around $J t = 22$ and $J t = 28$), formed by
the two wavefronts that, in this case, are not symmetrically evolving.   

\begin{figure}[h]
\centering
\includegraphics[width=0.50\columnwidth]{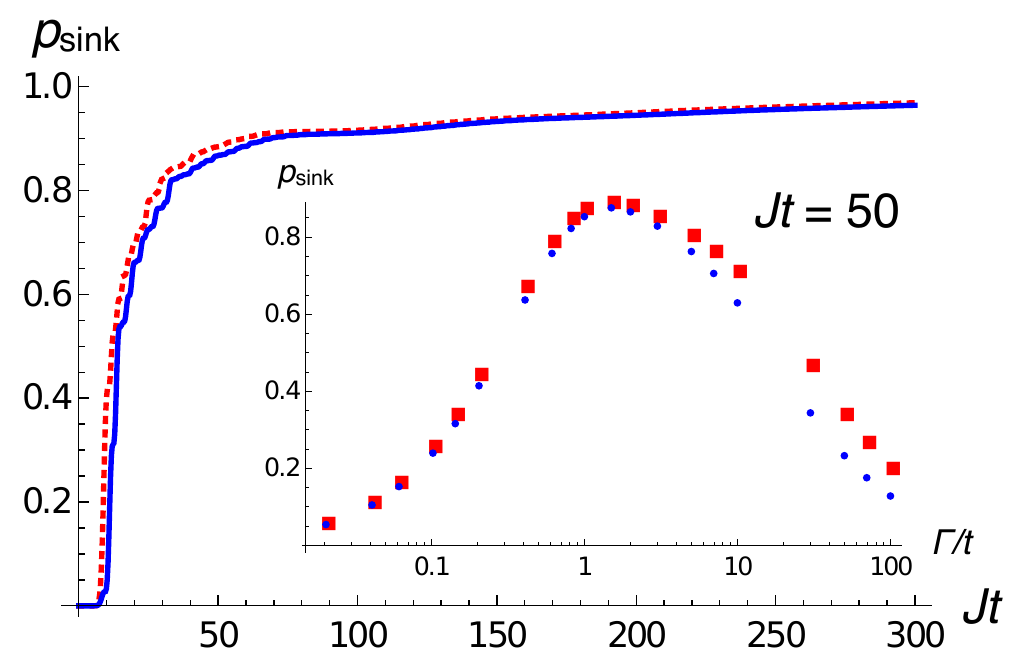}
\caption{The sink population $p_{\rm sink}(t)$ as a function of $J t$ for the isotropic (red, dashed line) and the
    dimerized (blue solid line) ring. Inset: the sink population at fixed time $J t = 50$  for the isotropic
(red squares) and the biomimetic (blue circles) configurations as a function of $\Gamma/J$. For both
configurations the optimal sink rate $\Gamma$ is achieved for $\Gamma/J \approx 2$.} 
\label{fig:gGammaOpt}
\end{figure}

In the case of the biomimetic configuration it is therefore not immediate to quantify
the transfer efficiency. Following Ref.~\citen{caruso09} we introduce an extra site, that we indicate by $\ket{s}$, that acts as a sink to 
which the population is irreversibly transferred from site $N/2+1$. Once defined the operator $S = \ketbra{s}{N/2+1}$, the transfer to the sink is 
modelled by a Lindblad term
\begin{align}
    \mathcal{L}_{\rm sink} [\rho] & = \Gamma \left [ 2 S \rho S^\dagger
    %\right . \nonumber \\
    %&\left .
    - \{S^\dagger S,\rho\} \right ].
    \label{eq:lindSink}
\end{align}
The efficiency of the transfer at a given time $t$ will be defined as the occupation probability of
the sink site, i.e.
\begin{equation}
    p_{\rm sink}(t) = \bra{s}\rho(t) \ket{s},
    \label{eq:eff}
\end{equation}
where $\rho(t)$ is the solution of the Lindblad master equation
\begin{eqnarray}
    \frac{d}{dt} \rho(t) = -i [H_{\rm R/LH},\rho(t)] + \mathcal{L}_{\rm sink}[\rho(t)],
         \label{eq:lindSinkME}
\end{eqnarray}
with initial condition $\rho(0)=\ketbra{1}{1}$. We now analyse the dependence of $p_{\rm sink}(t)$ on
the parameter $\Gamma$ looking for an optimal configuration. We remark that this measure is
equivalent to other transfer measures (see, for example Ref.~\citen{caruso09}).  As the inset of \fref{fig:gGammaOpt} shows, the
population of the sink at a time $J t = 50$, the transfer rate is optimal when it assumes values in a neighbourhood of $J$. 
As one may expect, for values of $\Gamma$ much smaller than the (average) coupling strength between the ring
sites, the transfer to the sink site much slower than the ring evolution; on the other side, for
$\Gamma$ larger than the excitation exchange average rate, the evolution of the site $j=17$ gets
``frozen'' because of a quantum-Zeno-like effect \cite{itano09}.  In what follows, we
therefore adopt the optimal value $\Gamma = 2 J$. In \fref{fig:gGammaOpt}
we show an example of the population of the sink as a function of time for $\Gamma/J=2$
for the isotropic and biomimetic ring configurations. We observe that the sink populations for the
two ring configurations are comparable, even though the dimerized ring shows a slightly reduced
transfer capability.

%\\[5pt]\adde{Nel lavoro con Alessia abbiamo considerato la popolazione del sink come misura della qualita` del trasferimento. Nei lavori seminali di Caruso, invece si usa la misura
%\[
%    \Gamma_{N+1} \int_0^t \rho_{N+1,N+1}(t') dt'
%\]
%Cerchiamo di capire le differenze. Magari la nostra misura e` piu` severa o meno anche se penso che
%le due misure siano monotone una rispetto all'altra.
%}
%
\section{The effect of disorder and dephasing} 
The models presented in the previous section are idealized. Any physical realization of the lattice
will in fact be subjected to different kinds of imperfections; for instance, the on site energies of the TLSs
composing the system might differ from each other or the interaction between nearest neighbour can vary
from site to site, e.g. because of different relative distances. Moreover, the system can be
affected by noise sources: if the TLSs are embedded in a scaffold, they will experience effects due,
e.g. to the vibrations (phonons) of the latter. The specific kind of disorder and noise influencing the
dynamics of the system, however, is very dependent on the particular physical realization. In the
case of the light harvesting complex LH1 our dimerized model is inspired by, even the mere structure
of the complex is still debated. \cite{scholes00,timpmann05,sener07} 

In this work we adopt a paradigmatic approach and consider static (i.e. time-independent) randomly distributed disorder
affecting only the on-site energies of the ring Hamiltonians \eref{eq:hamR} and \eref{eq:hamLH1}. Such disorder is
represented by a diagonal  term
\begin{align}
    H_{\rm dis}^\sigma(\vec{\mathcal{E}}) = \diag\llrr{\mathcal{E}_1, \mathcal{E}_2,\ldots,\mathcal{E}_N},
    \label{eq:disorder}
\end{align}
with $\mathcal{E}_i$ independent random variables with zero-mean Gaussian distribution with standard
deviation $\sigma$, quantifying the amount of disorder. Disorder introduces a random on-site energy
detuning between different sites of the ring which tends to localize the
eigenstates of the Hamiltonian. This results in a localization of the evolving state within a
typical $\sigma$-dependent localization length (Anderson localization \cite{abrahams79, malyshev04})
thus reducing the transfer capabilities of the ring \cite{tamascelli13}.

\begin{figure}[h]
%\subfigure[]{\label{fig:gDis} \includegraphics[width=0.45\columnwidth]{gSigma}}
%\subfigure[]{\label{fig:gDeph} \includegraphics[width=0.45\columnwidth]{gDeph}}
\begin{center}
\includegraphics[width=0.95\columnwidth]{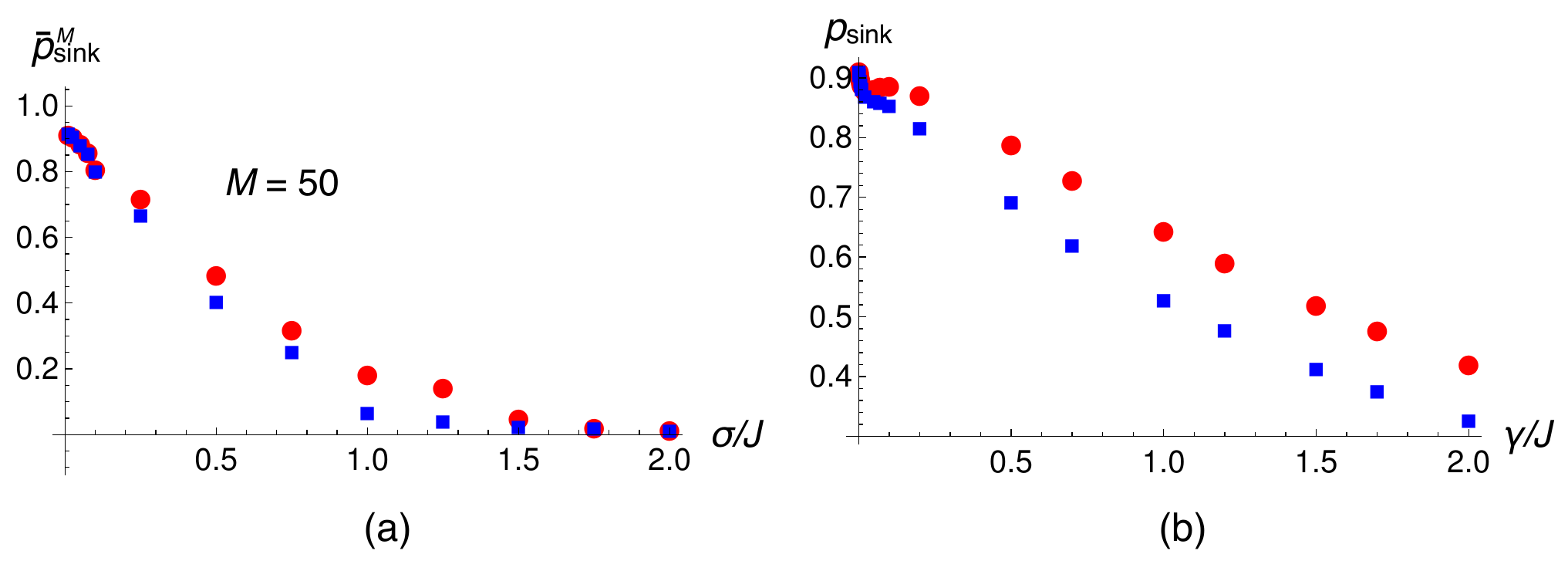}
\end{center}
\vspace{-0.3cm}
\caption{(a) The transfer efficiency $\bar{p}_{\rm sink}^M(t)$ as defined in \eref{eq:avTrans} at
time $J t = 100$ as a function of the ratio $\sigma/J$ for the isotropic and dimerized rings and $M=50$. (b) The transfer efficiency
\eref{eq:eff} in the presence of dephasing alone as a function of $\gamma/J$. In
both frames red circles refer to the isotropic ring whereas blue squares refer to the
biomimetic configuration.}
\label{fig:fDisDeph}
\end{figure}

We now quantify the effect of disorder on the transfer process for a fixed value of $\sigma$.
In order to make the results independent of the particular
realization of the diagonal term \eref{eq:disorder}, we need to simulate the evolution of the system for a large number $M$ of realizations
of the stochastic Hamiltonian part. For each realization $H_{\rm dis}(\vec{\varepsilon}_m)$ we numerically determine the  states
$\rho_m(t), \ m=1,2,\ldots,M$, as the solution of \eref{eq:me} with $\vec{\varepsilon} = \vec{\varepsilon}_m$.  
The transfer efficiency is defined as:
\begin{equation}
    \bar{p}_{\rm sink}^M(t) = \frac{1}{M} \sum_{m=1}^M \bra{s}\rho_m(t) \ket{s}.
    \label{eq:avTrans}
\end{equation}
In \fref{fig:fDisDeph}(a) we plot the sink population as a function of the ratio 
$\sigma/J$ for the isotropic and dimerized ring configurations at a time $J t = 100$. The
transfer efficiency, computed over an average of $M=50$ random realizations is a decreasing function of $\sigma$, as expected.
We point out that the transport over the dimerized ring
is much more affected by disorder than in the case of a ring with isotropic couplings.  

\begin{figure}[h]
%\subfigure[]{\label{fig:gDephWeak} \includegraphics[width=0.45
%\columnwidth]{gDephWeak}}
%\subfigure[]{\label{fig:gDephStrong}
%\includegraphics[width=0.45\columnwidth]{gDephStrong}}
\begin{center}
\includegraphics[width=0.95\columnwidth]{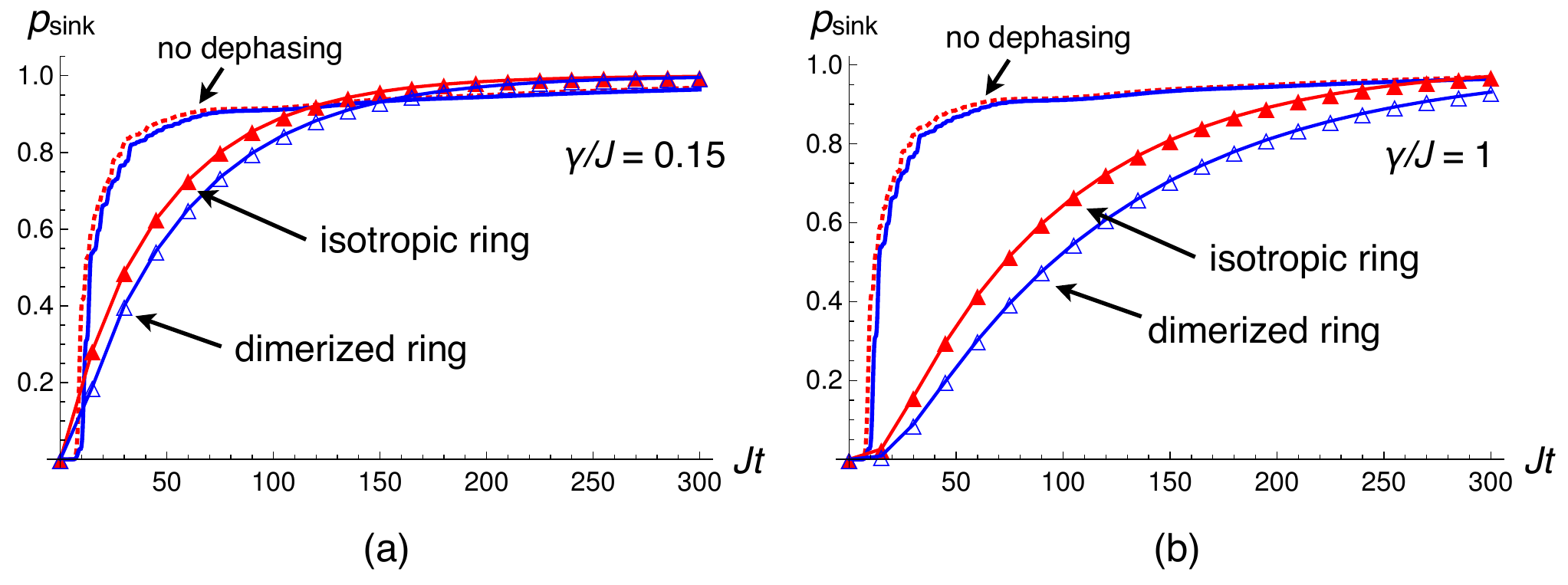}
\end{center}
\vspace{-0.3cm}
\caption{(a) The transfer efficiency $p_{\rm sink}$ in the presence of pure dephasing for the isotropic ring (red
    filled triangles) and the dimerized ring (blue empty triangles) as a function of $J t$ for (a) $\gamma/J=0.15$
    (b) $\gamma/J=1$. For comparison purposes we also report the population of the sink without dephasing (see figure~\fref{fig:gGammaOpt}).}
\label{fig:dephWeakStrong}
\end{figure}

The interaction of the system with a very large number of degrees of freedom, i.e. an
environment, can induce different effects to the system. As mentioned above the
vibration of the trapping lattice or, in the case of LH1, the proteic scaffold surrounding the TLSs,
however, can change their relative positions thus making the coupling coefficients time-dependent. In this
case the coupling coefficients would be functions of time. 
%that cannot be known without a detailed knowledge of the environmental degrees of freedom.
Another noise source is the interaction of each
TLS with its local environment, leading to fluctuations of its on-site energy. Such
fluctuations lead to a broadening of the line-shape spectrum of each TLS \cite{malyshev041,malyshev05,malyshev06}.
Following Ref.~\citen{chin10}, here we adopt a Markovian effective description of the interaction of each TLS with its local
environment and consider only pure dephasing local terms. Such effective description can be
seen as a coarse-graining of the stochastic random fluctuations of the
on site energies and results in the suppression of  the phase coherences of any superposition state of the system. The dephasing
process can be modelled by the Lindblad-form super-operator:
\begin{align}
    \mathcal{L}_{\rm deph}[\rho] &= \gamma \sum_{j=1}^N 2 \ketbra{j}{j} \rho \ketbra{j}{j} - \llrrb{\ketbra{j}{j},\rho},
    \label{eq:lindDeph}
\end{align}
where we considered the dephasing rates equal for all sites.
Figure \ref{fig:fDisDeph}(b) shows the transfer efficiency \eref{eq:eff} at the same time $J t = 100$
as in \fref{fig:fDisDeph}(a) for the isotropic and
dimerized rings in the absence of disorder ($\sigma=0$)  as a function of the ratio
$\gamma/J$. The loss of coherences detriments the transfer process by making it much slower.  It is
easy to show, moreover, that in the
limit $\gamma \gg J$ the excitation spreads on the lattice at a rate that is
close to a purely diffusive process. However, for long times pure dephasing leads to a higher
population of the sink site w.r.t. the purely coherent evolution. In \fref{fig:dephWeakStrong} we
show an example of the effect of dephasing for both a small and a large value of the ratio $\gamma/J$. We observe that dephasing affects the excitation transfer to the
sink more in the case of a biomimetic ring configuration than in the case of a isotropic ring and
such difference becomes more pronounced for larger values of $\gamma$ w.r.t. $J$.

\section{Dephasing assisted transport}
As shown in the previous section, disorder and dephasing both decrease the transfer capability over a
circular graph.  The physical mechanisms behind such efficiency reduction, however, are
completely different. On the one hand, disorder induces random phases in the state of the system, that
lead to the destructive interference that inhibits the spreading of the excitation over the
lattice; on the other, pure dephasing destroys the phase relations between the different sites of the same
lattice that determine super-diffusive propagation of the excitation typical of quantum walks
\cite{mulken11}. In this section we show that
dephasing can indeed enhance the transfer efficiency in the presence of disorder and we address a characterization
of DAT over the two ring models we are considering.

\begin{figure}[h]
\begin{center}
\includegraphics[width=0.95\columnwidth]{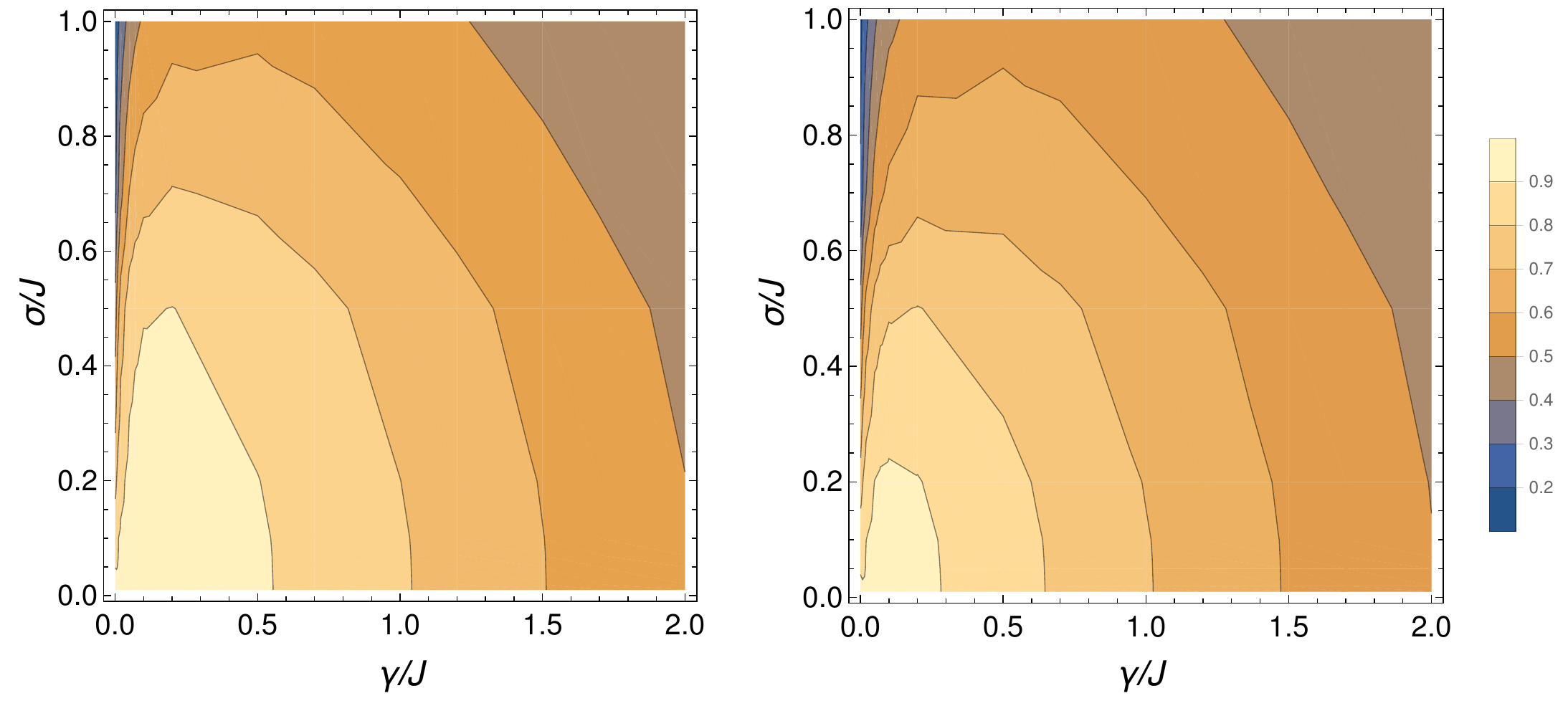}
\end{center}
\vspace{-0.3cm}
\caption{Contour-plots of the transfer efficiency $\bar{p}_{\rm sink}^M(t)$  \eref{eq:avTrans} at
time $J t = 100$ as a function of the ratio $\gamma/J$ and of 
$\sigma/J$ for (a) the isotropic ring and (b) the dimerized ring.
}
\label{fig:fDisANDDeph}
\end{figure}

\begin{figure}[h]
\centering
\includegraphics[width=0.50\columnwidth]{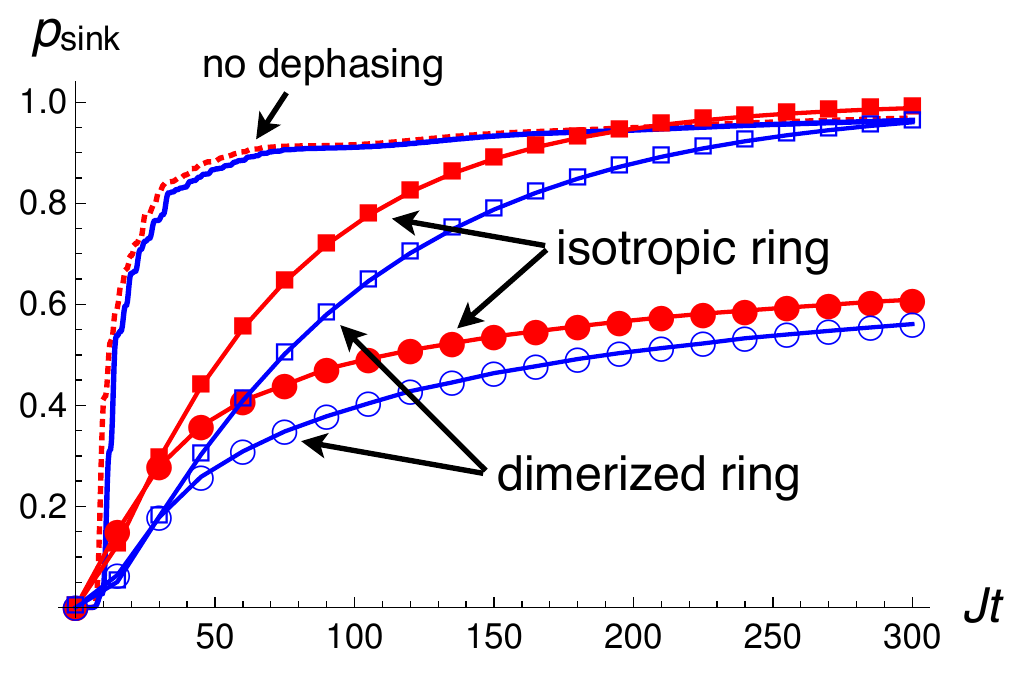}
    \caption{The (average) sink population as a function of $J t$. Circles: sink population in the presence of disorder with $\sigma/J=0.5$.
    Squares: sink population in the presence of the same amount of disorder and dephasing for $\gamma/J=0.1$ Filled red markers refer to the isotropic case, empty blue markers to the 
    dimerized configuration. Dashed red line and solid blue line represent the sink population in the
    absence of dephasing and disorder for the isotropic and dimerized ring respectively and are reported for comparison purposes.} 
\label{fig:gDisDephOpt}
\end{figure}

The complete master equation determining the evolution of the density matrix of the system is now:
\begin{align}
    \frac{d}{dt} \rho(t) = -i [H_{\rm R/LH1} + H_{\rm dis}^\sigma(\vec{\mathcal{E}}),\rho(t)]+
    \mathcal{L}_{\rm deph}[\rho(t)] + \mathcal{L}_{\rm sink}[\rho(t)].
    \label{eq:me}
\end{align}
We will consider an initial condition $\rho(0)  =\ketbra{1}{1}$ as in the previous sections
and study the behaviour of the sample transfer probability \eref{eq:avTrans} for different ratios $\sigma/J$ and  $\gamma/J$. Figure
\ref{fig:fDisANDDeph} shows the $\bar{p}_{\rm sink}^M(t)$ for $M=50$ and $J t=100$. It is evident
that for both the isotropic and dimerized rings, for any fixed value of $\sigma/J$ there is a
range of values $\gamma/J$ that lead to an improved transfer capabilities.  For
any choice of $\sigma/J$ and $\gamma/J$, moreover, the transfer over the isotropic ring is
always more efficient than the transfer over the dimerized configuration. 
An intuitive explanation of can be given as follows: disorder introduces random phases between
adjacent sites that would be responsible of the localization of the excitation in the neighbourhood
of its initial condition, while dephasing tends to ``wash-away'' the ensuing interference
patterns. In \fref{fig:gDisDephOpt} we show a comparison between the sink
population in the presence of the sole disorder ($\sigma/J=0.5$) and in the presence of the same
amount of disorder and dephasing ($\gamma/J=0.1$). The selected parametrization is suggested by an
inspection of \fref{fig:fDisANDDeph}.
Dephasing, therefore, enhances transport in the presence of disorder already at a time-scale
comparable to the propagation time of the excitation from its initial location to the opposite site
$j=17$. The suppression of the localization of the excitation is made more evident by the
faster increase of the sink population. Even in the presence of disorder, moreover, dephasing still
allows, in the long-time limit, to increase the sink population beyond the value reached
in the absence of both dephasing and disorder.

\section{Conclusion and Outlook}
Dephasing assisted transport is an example of how an interplay between the coherent and
incoherent part of system
dynamics can lead to an improvement of the energy transfer between different points of graph. In
this work we have investigated DAT over a circular graph with isotropic and
alternate couplings between nearest-neighbour sites in the presence of disorder and
dephasing. Our analysis confirmed the existence of regimes where DAT occurs.

We showed that dimerization does not
provide an advantage for the energy transfer process over disordered systems affected by
decoherence. While the model of the system and of the effects induced by the
system-enviromnent interaction we adopted is highly simplified and might not capture some key
feature of the light-harvesting complexes, our results hint that the functional advantage of the dimerized
structure of LH1/2 complexes cannot be captured by a quantum-walk perspective.

Future work will address the inclusion of losses, dissipation and thermalization, as
to understand whether the time-scales on which DAT occurs are compatible with the exciton loss and
relaxation rates of the system.  More refined models \cite{malyshev041,malyshev05,malyshev06}, where the effective Markovian
dynamics is replaced by scattering processes on the lattice hosting the TLSs will be
considered as well.

%\bibliographystyle{ws-ijqi}
%\bibliography{biblio} 

\end{document}